\documentclass{article}
\usepackage{epsfig}
\begin{document}
\title{SPONTANEOUSLY
BROKEN CHIRAL SYMMETRY AND HARD QCD PHENOMENA}
\author{M.J. Amarian, V.M. Braun, S.J. Brodsky, J. Collins, M. Diehl,\\ G. Ecker, A.V. Efremov, T. Feldmann, L.L. Frankfurt, K. Goeke,\\ D.S. Hwang, I.B. Khriplovich, L. Mankiewicz, G.A. Miller,\\ P.J. Mulders, M. Musakhanov, J.W. Negele, N.N. Nikolaev,\\ K. Passek-Kumericki, V. Petrov, P.V. Pobylitsa, M.V. Polyakov,\\ M. Praszalowicz, A.V. Radyushkin, E.V. Shuryak, P. Stoler,\\ M. Strikman, M. Vanderhaeghen, T. Walcher, W. Weise, C. Weiss}
\date{Physikzentrum Bad Honnef, Germany\\ July 15-19 2002}

 \maketitle
 \begin{center}
 ABSTRACT
 \end{center}

 These are the mini-proceedings of the workshop ``Spontaneously Broken Chiral Symmetry and Hard QCD Phenomena" held
 at the Physikzentrum Bad Honnef from July 15 to 19, 2002. 
Every author presents a summary of his talk. The transparencies of all speakers and further information on the workshop can be found on the website:  

\begin{center}
 \textbf{http://www.tp2.ruhr-uni-bochum.de/hardreactions.html}
\end{center}

\section*{PREFACE}

\vspace{0.5cm}
Quantum chromodynamics (QCD), being a non-abelian gauge theory, is
considered now as the fundamental theory of strong interactions. Experiments
at high energy accelerators: SLAC (Stanford), CERN (Geneva), DESY (Hamburg),
FNAL (Chicago) found elementary building blocks --- i.e. quarks and gluons
--- and checked that the interaction between them is well described
by the equations of QCD. Numerous experiments confirmed that the interaction
between elementary particles is decreasing with increase of resolution.
This feature corresponds to a decrease of the space-time interval
probed in the process and requires a high momentum transfer to the
elementary particles (hard process). Thus the Lagrangean of QCD seems
to be established both theoretically and experimentally. Challenging
problems now are in search for novel QCD phenomena in the case of
large hadron (quark-gluon) densities, where the smallness of the coupling
constant is insufficient to support applicability of perturbative
approaches, and in the case of low energy phenomena in hadronic and
nuclear physics where the coupling constant is not small.

Thus one challenge is to build theoretical methods of solving equations
of QCD in the regime where effective interaction is not small any
more. These problems are becoming more acute with time because various
developements suggest the possible existence of plenty of new QCD phenomena,
as there are recent observed astrophysical phenomena in neutron stars,
strong indications from high energy accelerators like the fast increase
of gluon densities with energy, or new theoretical ideas on the possibility
of variety of phase transitions in hadronic matter.

The phenomenon of spontaneously broken chiral symmetry (SBCS) together
with confinement of color is usually considered as the most important
QCD properties and the cornerstones for the understanding of the physics
of strong interactions. It is generally believed that the existence of
nuclear physics phenomena (and therefore of life!) is closely related
to the spontaneous chiral symmetry breaking. So a
lot of researchers suggested a variety of ideas how to describe this
phenomenon. These approaches are all based on chiral dynamics of the
QCD Lagrangean. Many of the predictions were checked experimentally
at various accelerators. Appropriate phenomena are investigated now
e.g. at SIN (Switzerland), MAMI (Mainz) and ELSA (Bonn). In spite
of the evident success in the explanation of a variety of existing
data such an approach has the problem to explain the mass of the $\eta^{\prime} $
meson directly from the QCD Lagrangian. Interesting ideas how to solve
this puzzle are known for a long time but only now they become of
practical interest because of the possibility to study this problem
experimentally. COMPASS (CERN) investigates now phenomena related to
the corresponding QCD anomaly. The low energy community accumulated
so far a vast amount of interesting phenomenological data and interesting
ideas how to explain them. Still a success of low energy approaches
has limitations because of important roles of virtual high energy
phenomena. At the same time subtraction constants introduced to account
for them can be investigated directly in the high energy phenomena.
This physics is natural for intermediate energy machines at JLab or
DESY (HERMES). Thus overlap between interests of researchers working
on low energy and high energy phenomena in the field of spontaneously
broken chiral symmetry is evident.

For many years the theoretical methods used by high- and low-energy
communities were essentially different. In particular, the traces
of chiral QCD phenomena were not very pronounced. However, the situation
changed few years ago when it was noticed that a combination of QCD
factorization theorems with chiral algebra can lead to important consequences
that can be tested experimentally. Indeed, quite often the quantities
measured in high energy processes can be factorized into a \char`\"{}hard\char`\"{}
part, describable mainly in terms of perturbative QCD-methods, and
a nonperturbative \char`\"{}soft\char`\"{} part. It is the soft part
that can be analyzed in terms of sponataneously broken chiral symmetry.
This conference is an attempt to give the possibility for low energy
and high energy communities to exchange information, ideas and to
overcome the gap between them.

The basic object of spontaneously broken chiral symmetry --- namely
the wave function of the pion being a pseudogoldstone mode --- is now
investigated in the electromagnetic pion form factor at JLab, in production
of pions and $\eta$ in two photon processes at LEP, or in diffractive
production of dijets at FNAL. All experiments produce consistent results.
Hard exclusive electroproduction of vector and pseudoscalar mesons
and deeply virtual Compton scattering, owing to the QCD factorization
theorems, allow us to access the structure of hadrons in a new theoretically
well controlled way. These processes attracted great interest in both
communities and, as a consequence, the communication between them
has intensified.

The objective of the present workshop was to invite experts from both
groups, who expressed their interest in combining chiral symmetry
with hard processes. We choose basically two domains of applications
of chiral QCD dynamics:

1. Chiral symmetry at low energies.

2. Chiral symmetry at high energies.

It was agreed not to publish printed proceedings of the workshop but
rather to ask the participants to present a one-page summary of their
contributions with references to the original papers. These contributions
are presented below. The transparencies of the talks presented at
the workshop can be found on the website

\begin{center}
\textbf{http://www.tp2.ruhr-uni-bochum.de/hardreactions.html}
\vspace{.8cm}

\sc Leonid Frankfurt, Klaus Goeke, Pavel V. Pobylitsa,\\
 Maxim V.~Polyakov, Mark Strikman, Wolfram Weise

\end{center}

\begin{center}
\section*{New Results on DVCS and \\ Exclusive Meson Production at Hermes}
M.J.~Amarian
\end{center}
\maketitle


\begin{center}
(on behlf of the HERMES Collaboration)\\
DESY, Notkestrasse 85, D - 22603 Hamburg, Germany
\end{center}


We present experimental results on beam-spin asymmetry (BSA)
in Deeply Virtual Compton Scattering (DVCS)  using new 
set of data taken in year 2000 on hydrogen target. 
Extracted value of the asymmetry is consistent with
published results \cite{ref1} based on 1996/97 data 
taking period. In addition we present preliminary results
on first measurement of lepton-charge asymmetry (LCA) in DVCS on
hydorgen target using electron and positron beams of HERA \cite{ref2}. 
While the BSA asymmetry gives an access to imaginary part of DVCS
amplitude, the LCA gives an access to real part of DVCS amplitude.
Comparison of the sign and the magnitude of LCA with existing models
gives a constrain on the so called D-term of polinomiality condition
of Generalized Parton Distribution \cite{ref3}. 

Second subject is devoted to single target-spin asymmetry in
electroproduction of pseudoscalar mesons. Here we present new
experimental results based on polarized deuteron data. 
Extracted value of single-spin asymmetry of semi-inclusive 
positive pions is diluted compared to the published hydrogen 
target data \cite{ref4}. When plotted versus z (the ratio of the
energy of produced pion over virtual photon energy) for both
exclusive $\pi^+$ and $\pi^-$ asymmetry rizes dramatically
and has the same sign, being consistent with published data
on hydrogen target for $\pi^+$ \cite{ref5}. The sign of asymmetry
for netral pions at high z is opposite to the charged ones. 
In addition we present for the first time experimental results 
on single-spin asymmetry for identified kaons.



\begin{center}
\section*{Coherent dijet production in QCD}
{V.M.~Braun}${}^1$,
{S.~Gottwald}${}^1$,
{D.Yu.~Ivanov}${}^{1,2}$
{A.~Sch\"afer}${}^1$ and
{L.~Szymanowski}${}^{3,4}$\end{center}


\begin{center}${}^1${
 Institut f\"ur Theoretische Physik, Universit\"at
   Regensburg, \\ D-93040 Regensburg, Germany
                       } \\[0.2cm]
\vspace*{0.1cm} ${}^2$ {
Institute of Mathematics, 630090 Novosibirsk, Russia
                       } \\[0.2cm]
\vspace*{0.1cm} ${}^3${
    CPhT, \'Ecole Polytechnique, F-91128 Palaiseau, 
    France\footnote{Unit{\'e} mixte C7644 du CNRS.}
                       } \\[0.2cm]
\vspace*{0.1cm} ${}^4$ {
 Soltan Institute for Nuclear Studies,
Hoza 69,\\ 00-681 Warsaw, Poland
                       }

 \end{center}

  Coherent production of hard dijets by incident pions and/or linearly 
 polarized real photons can provide direct measurement for the pion 
 distribution amplitude, clear evidence for chirality violation in hard 
 processes, the first measurement of the magnetic susceptibility of the 
 quark condensate and the photon distribution amplitude.    
 It can also serve as a sensitive probe of the generalized gluon parton 
 distribution.

  We consider the possibility to calculate  the 
  coherent dijet production cross section using QCD collinear factorization.
  For the pion, we find that the collinear factorization is valid in the 
  double 
  logarithmic approximation but does not hold beyond because of soft 
  gluon exchanges in the so-called Glauber region.   
  For the real photon the collinear factorization is valid in 
  perturbation theory, but is violated in the 
  chirality-breaking contribution corresponding to the nonperturbative 
  photon wave function. Detailed calculations are presented for the 
  FERMILAB and HERA kinematic regions.


\begin{center}
\section*{Light-Front Wavefunctions and Hard Exclusive Processes in QCD \footnote{Research
  partially supported by the Department of Energy under contract
  DE-AC03-76SF00515}}
S.J.~Brodsky
\end{center}
\maketitle

\begin{center}
Stanford Linear Accelerator Center \\
2575 Sand Hill Road \\
Menlo Park, CA 94025, USA\\
E-mail: sjbth@slac.stanford.edu\end{center}

Hard hadronic exclusive processes are now at
forefront of QCD studies, particularly because of their role in
illuminating hadron structure at the amplitude level and in
interpreting the physics of exclusive heavy hadron
decays.
In the
accompanying transparencies, I discuss some of the
theoretical issues and empirical challenges to the application of
perturbative QCD and factorization theorems, such as the role of soft and
higher twist QCD mechanisms, the role of relative orbital angular
momentum, single-spin asymmetries in deeply virtual Compton scattering,
the behavior of the ratio of Pauli and Dirac nucleon form factors, the
apparent breaking of color transparency in quasi-elastic proton-proton
scattering, and the measurements of hadron and photon wavefunctions in
diffractive dijet production.
For recent reviews, see ref.~\cite{Brodsky:1989pv,Brodsky:2002st}

Electromagnetic, electroweak, and gravitational form
factors~\cite{Brodsky:1980zm,Brodsky:1998hn,Brodsky:2000ii} and
the deeply virtual Compton amplitude~\cite{Brodsky:2000xy,Diehl:2000xz} have an
exact representation as overlap integrals of light-front wavefunctions.
The Pauli Form factor and the
$E$ generalized parton distribution require overlaps of wavefunctions
differing by one unit of relative orbital angular momentum
projection $L_z$.  The gravitational spin-flip amplitude  $B(q^2)$
vanishes at $q^2=0$ in agreement with the equivalence
principle.~\cite{Brodsky:2000ii}  In the case of timelike form factors
such as
$B\to M\ell \nu$, and DVCS~\cite{Brodsky:2000xy,Diehl:2000xz}, off-diagonal
$\Delta n = 2$ contributions are required.~\cite{Brodsky:1998hn}  These exact
light-front representations provide the starting point for the PQCD treatment
of form factors and other hard scattering amplitudes.  It should be emphasized
that the leading-twist contributions to exclusive hard processes are by
definition devoid of any integration region where the parton is close to the
mass shell.  I also note that any model of a soft wavefunction which has
Gaussian fall-off in the transverse directions, must also have Gaussian
fall-off at the $x=0$ and $x \to 1$ endpoints.  It is thus inconsistent to fit
the $x$ distributions of such models to the empirical power-law $(1-x)^n$
forms, since the power law  behavior is accounted for by perturbative
QCD. ~\cite{Brodsky:1994kg}

Perturbative QCD and its
factorization properties at high momentum transfer provide a critical
guide to the phenomenology of exclusive amplitudes.~\cite{Brodsky:1989pv} The
leading-twist PQCD predictions have many empirical successes.  The reduced
amplitude formalism allows the application of perturbative QCD to hard
nuclear amplitudes.~\cite{Brodsky:2001qm}  The fact that the hadron
enters through its distribution amplitude---its valence wavefunction
evaluated at impact parameter separation $b \sim 1/Q$---leads to the color
transparency of quasi-elastic hard-scattering in nuclei and diminished
initial and final state hadron interactions.~\cite{Brodsky:xz}

The chiral structure of hard scattering amplitudes, and the
fact that the distribution amplitudes are $L_z=0$ projections of the
hadron light-front wavefunctions lead to hadron-helicity conservation at
leading twist.~\cite{Brodsky:1981kj}  Hiller, Hwang, and
I~\cite{BHH,Brodsky:2002st} find that a  PQCD-motivated
form~\cite{Lepage:1980fj} for the ratio of Pauli and Dirac nucleon form factors
${F_2(Q^2)\over F_1(Q^2)} \sim {\log^\gamma Q^2\over Q^2}$  is consistent with
recent data from Jefferson Laboratory.  The normalization discrepancy between
measurements of the spacelike pion form factor and PQCD predictions may be due
to a systematic error in extrapolating electroproduction to the pion pole at
$t \to  m^2_\pi.$ The effective $(q
\bar q)(t) \to \pi$ form factor measured in electroproduction
$F(t,Q^2)$ is normalized to $\sqrt(-t) f_\pi$ not $f_\pi^2.$

I also discuss the
unexpected role of final state interactions in deep inelastic scattering,
and their role in shadowing theory,~\cite{Brodsky:2002ue} single-spin
asymmetries,~\cite{Brodsky:2002rv} and diffraction.  Similar corrections
to the quark propagator can modify DVCS predictions, even in
light-cone gauge.  I also review features of the model for leading-twist DVCS
amplitudes and Bethe-Heitler interference contributions developed by Close and
Gunion and myself~\cite{Brodsky:1971zh} which
incorporates analyticity, crossing behavior, Regge exchange, fixed Regge poles,
and leading-twist scaling.


\begin{center}
\section*{Factorization for hard exclusive processes}
John Collins
\end{center}


\begin{center}
        Physics Department,
        Penn State University, 
        104 Davey Laboratory,\\
        University Park PA 16802
        U.S.A.
\end{center}


A review was given of the rationale for factorization in processes
such as exclusive deep-inelastic production of mesons.

In high-energy collisions, large ranges of virtualities and rapidities
can be accessed, and a factorization theorem separates the different
scales and rapidities.  The asymptotic freedom of QCD enables
perturbative predictions to be made for short-distance coefficient
functions and evolution kernels.  Connections with non-perturbative
physics are made with the aid of operator definitions of (generalized)
parton densities and light-front distribution amplitudes.

First, leading regions in momentum space are identified, and they
lead to operator definitions of parton densities etc.  They also lead
to a coordinate-space interpretation that suggests that the results
should be valid beyond perturbation theory.  A systematic method of
successive approximation leads to a perturbative construction of hard
scattering coefficients and evolution kernels with subtractions to
avoid double counting.

There are a number of complications to overcome: (a) In the parton
densities, etc., there are divergences at large transverse momentum and
at large gluon rapidity; these must be suitably cutoff, renormalized,
or cancelled to obtain correct results.  (b) Gluon interactions must
be correctly treated so that gauge-invariant results are obtained.
(c) Soft final-state interactions must be shown to cancel (or in other
situations to factorize); the arguments involve the application of
contour deformation in momentum space and the use of the QCD Ward
identities, with substantial components of the argument going beyond
perturbation theory.

Generalizations of the arguments apply to other processes.  The main
ideas in the form used here can be found in \cite{DIS.Meson,DIS.diff}
and the references quoted there.



\begin{center}
\section*{Generalized Parton Distributions \\ in Transverse Space}
M.~Diehl
\end{center}
\maketitle


\begin{center}
Institut f\"ur Theoretische Physik E,
RWTH Aachen, 52056 Aachen, Germany
\end{center}


The impact parameter formulation allows for a simple physical
interpretation of quantities like parton distributions and form
factors in terms of quarks and gluons.  I investigate the specifics of
this representation for generalized parton distributions at nonzero
skewness $\xi$ and their relation with Lorentz symmetry on the light
cone.  An explicit realization of this representation is obtained when
the parton distributions are expressed in terms of hadronic light-cone
wave functions.

The primary aim of this representation is to provide a framework to
investigate exactly how generalized parton distributions
simultaneously describe longitudinal and transverse structure of a
fast moving hadron, and to compare this information with the one in
elastic form factors, in ordinary and in $k_T$ dependent parton
distributions.

Details on the material presented in this talk can be found in
\cite{Diehl:2002he}. This study builds on work by Burkardt for the
case of zero skewness $\xi=0$ \cite{Burkardt:2000za}, and is closely
related with ideas recently formulated in \cite{Ralston:2001xs}.



\begin{center}
\section*{Recent developments \\ in chiral perturbation theory}
G. Ecker
\end{center}
\begin{center}
Inst. Theor. Physik, Univ. Wien, A-1090 Wien, Austria
\end{center}
\vspace*{.5cm} 
After a short introduction to chiral perturbation theory, the
effective field theory of the standard model at low energies, the
present status of the field is briefly reviewed.

Recent advances in the chiral analysis of pion pion 
scattering \cite{cgl} are discussed in some detail. Combining the 
chiral amplitude of next-to-next-to-leading order in the low-energy 
expansion with experimental input at higher energies via
dispersion relations (Roy equations), threshold parameters and phase
shifts up to cms energies of 800 MeV have been determined with
unprecedented precision. Comparison with recent
$K_{e4}$ data from the BNL-E865 experiment corroborates the status of
the quark condensate as dominant order parameter of spontaneous chiral 
symmetry breaking. As a consequence, we have now direct experimental
evidence that more than 94 $\%$ of the pion mass is due to the leading
term in the chiral expansion. 

In general, matching between the chiral low-energy structure and the
high-energy behaviour of amplitudes requires nonperturbative
methods to bridge the gap between the two regions. The chiral
resonance Lagrangian is a phenomenologically successful example where 
the matching produces the correct low-energy constants of 
next-to-leading order.

Four-pion production in $e^+ e^-$ annihilation and in $\tau$ decays is 
a case where the leading resonance couplings are not sufficient to
extend the low-energy amplitudes into the resonance region. 
Surprisingly, the amplitudes in the chiral limit available
in the literature were found to have the wrong normalization.
The next-to-leading-order calculation allows for a straightforward
continuation into the resonance region. Unlike for the pion form
factor, for instance, additional ingredients are necessary for such an 
extrapolation like $\omega$ and $a_1$ exchange that only show up 
in higher orders of the low-energy expansion. With the present form 
of the amplitudes \cite{eu02}, the $e^+ e^-$ cross sections are well 
described over a range of several orders of magnitude up to
about 1 GeV cms energy. Additional information about the high-energy 
structure of the amplitudes must be implemented before the full phase 
space available in $\tau$ decays can be covered.



\begin{center}
\section*{Collins analyzing power \\  and azimuthal asymmetries}
\underline{A.V.~Efremov}$^a$, K.~Goeke$^b$ and P.~Schweitzer$^c$
\end{center}
\maketitle

\begin{center}
{\footnotesize
$^a$ Joint Institute for Nuclear Research, Dubna, 141980 Russia\\
\footnotesize
$^b$ Institute for Theoretical Physics II, Ruhr University Bochum, Germany\\
\footnotesize
$^c$ Dipartimento di Fisica Nucleare e Teorica, Universit\`a di Pavia, Italy}
\end{center}

Spin azimuthal asymmetries in pion electro-production in deep
inelastic scattering off longitudinally polarized protons, measured by
HERMES, are well reproduced theoretically with no adjustable
parameters using the chiral quark soliton model for transversity
distribution and DELPHI result for averaged over $z$ Collins analysing
power $|{\langle H_1^\perp\rangle}/{\langle D_1\rangle}|=12.5\pm1.4\%$.

Predictions for azimuthal
asymmetries for a longitudinally polarized deuteron target are given.
>From $z$-dependence of the asymmetry and chiral quark soliton model for
transversity the z-dependence of the Collins
fragmentation function is extracted (see Fig.1).
\vspace{-5mm}
\begin{figure}[h!]
\begin{center}
\begin{tabular}{ll}
  \hspace{-0.8cm}
  \epsfig{figure
  =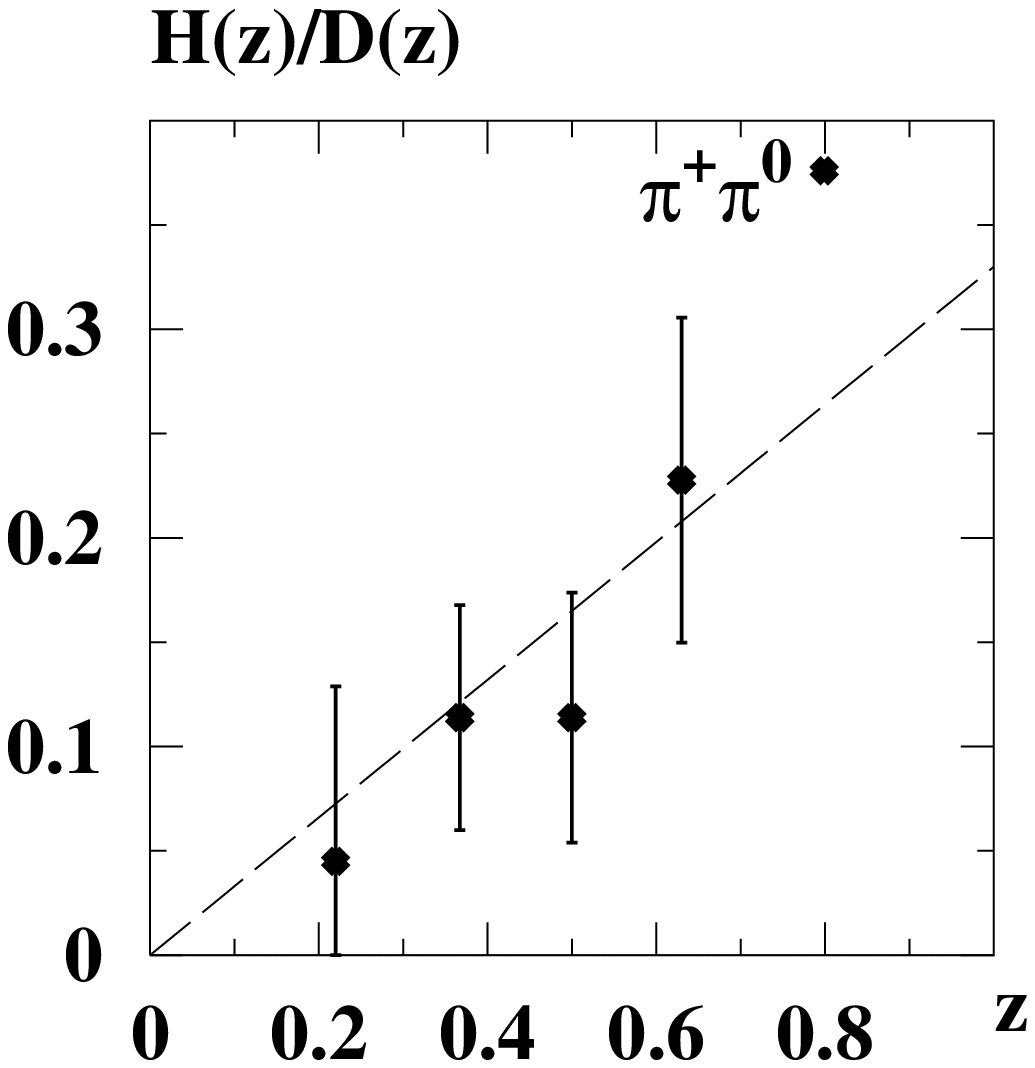,width=6.5cm,height=5.5cm}
    &
  \raisebox{0.8ex}{\mbox{\epsfig{figure=
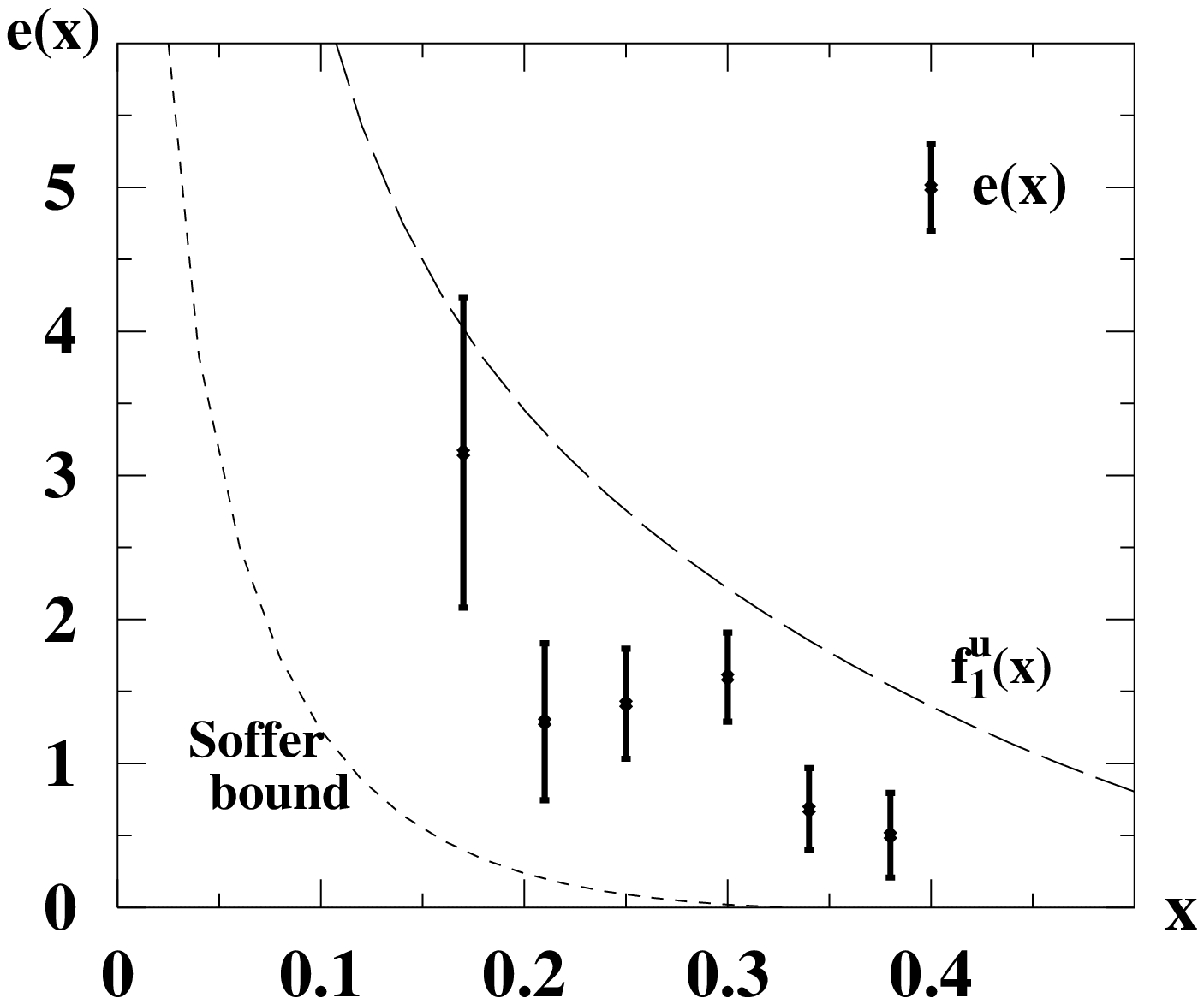,width=6cm,height=5.2cm}}}
    \\
  \begin{minipage}{6cm}{\footnotesize
  \vspace{-9mm}
  Fig.1. The Collins analyzing power
  $H_1^\perp/D_1$ vs. $z$, as extracted from HERMES data
  for $\pi^+$ and $\pi^0$ production.}
  \end{minipage}
    &
  \begin{minipage}{6cm}{\footnotesize Fig.2.
  The flavour combination
    $e(x)\!=\!(e^u\!+\!\frac{1}{4}e^{\bar d})(x)$,
    with error bars due to statistical error of CLAS data,
    vs. $x$ at $\langle Q^2\rangle \!=\! 1.5\,{\rm GeV}^2$.
    For comparison $f_1^u(x)$ and the twist-3 Soffer low bound
    are shown.}
    \end{minipage}
\end{tabular}
\end{center}
\end{figure}

\vspace{-4mm}
Using Collins analysing power the first information on $e(x)$ is
extracted (see Fig.2) from $A_{LU}^{\sin\phi}$ asymmetry of CLAS
collaboration. For more details and references see hep-ph/0206267.







\begin{center}
\section*{$\eta$ and $\eta'$ Mesons in Hard Exclusive
  Reactions\footnote{Talk
  available at http://tfeldman.home.cern.ch/tfeldman/badhonnef02.pdf}}
Thorsten Feldmann
\end{center}


\begin{center}
\it RWTH Aachen, Institut f\"ur Theoretische Physik~E, 52056 Aachen, Germany
\end{center}


Properties of $\eta$ and $\eta'$
mesons can be studied in various hard exclusive reactions:
The overall normalization of leading-twist light-cone distribution 
amplitudes (DAs)
that enter perturbative QCD calculations in the factorization approach
\cite{ERBL} is provided by weak decay constants. 
In general these are parametrized in terms of 
{\em two}\/ mixing angles that can be determined from theoretical
and phenomenological considerations \cite{KLTF}. 
In the quark-flavor basis the two mixing angles nearly coincide with 
a value of about $39^\circ$, indicating that OZI-rule violating
contributions to the {\em decay constants}\/ are small.
The shape of the DAs can be constrained by considering the 
$\eta\gamma$ and $\eta'\gamma$ transition form factors
at large momentum transfer \cite{Pgamma}. 
We found that pions, $\eta$ and $\eta'$ 
mesons have leading-twist 
DAs that are consistent with the asymptotic form already at 
low factorization scales. Furthermore, experimental data
excludes large ``intrinsic'' (constituent) charm in $\eta$ and
$\eta'$ mesons, in accordance with Refs.~\cite{charm}.
As a consequence, neglecting all OZI-rule violating
effects, simple ``Brodsky-Lepage'' \cite{BLtype} 
interpolation formulas for  $\eta(\eta')\gamma$ form
factors can be obtained, that are in appoximate agreement with
experiment for large {\em and}\/ small momentum transfer \cite{interpol}.
Two-body decays of heavy charmonia into $\eta/\eta'$ mesons
are sensitive to $\eta$\/--$\eta'$ mixing parameters, too
\cite{jpsi}. 
Particularly interesting are the decays of $J/\psi$ or $\psi'$
into $\eta(\eta')\omega$ which are related to the
``$\rho$\/--$\pi$\/--puzzle'' (see e.g.\ \cite{rhopi}
and references therein). 
A good knowledge of $\eta$ and $\eta'$ DAs and mixing
parameters is also
required to understand non-leptonic $B$ decays, in particular the
rather large $B \to \eta'K$ branching ratio (see \cite{martin}
and references therein.)





\begin{center}
\section*{Hard diffractive processes and light cone pion wave function.}
L.L.Frankfurt
\end{center}


\begin{center}
 Tel Aviv University, Israel
\end{center}

 
In the leading order over $\alpha_s$ the cross section of diffractive 
production of two high $\kappa_t$ jets unambigously follows from QCD 
factorization theorem ref.\cite{ref11,ref22} : as 
$\sigma(\pi +A\rightarrow 2jet +A)=
c{\alpha_s xG_A \Delta \psi_{\pi}(z,\kappa_t)}^2 {1/R_A^2}$. 
Here $\kappa_t$ is transverse momentum of jet, $\Delta$ is Laplace operator 
in 2 dimensional $\kappa$ space, $G_A$ is the gluon distribution in a nuclear 
target,$x={4k_t^2\over s}$,  $R_A$ is the radius of the nuclear target, c is 
the constant calculated in QCD. (For the review of the history of this 
process see ref.\cite{ref22}.) For the small x processes the major difference 
between skewed and usual gluon distributions is accounted for by QCD evolution
equation for the skewed parton distributions. At extreemly small x 
perturbative interaction becomes strong leading to significant absorbtion.

$A$ dependence observed in FNAL  ref.\cite{ref33} is faster than the 
prediction of ref.\cite{ref11} for the process where final nucleus 
is in the ground state. However experimental signature of the process is 
the strong forward peak $\propto F_A(t)^2$ so final nucleus may appear 
in excited state. Summing over nuclear excitations leads to the additional 
$A$ dependence of cross section which agrees well  with FNAL data.
(ref.\cite{ref22}  and this talk.) 

HERA data on hard diffractive processes
$\gamma^*+p\rightarrow V+p$ where $V=\gamma,\rho,\omega,\phi,\psi$.
demonstrated that vector mesons are squized by external hard 
probe.( H.Abramowicz talk at this conference). But just this property is 
relevant for the applicability of QCD factorization theorem.

To summarize: FNAL,HERA data confirmed predicted by QCD unusual properties
of hard diffractive processes and it is worth to use these processes to
investigate wave functions of hadrons. This talk is restricted by the
discussion of the process:$\pi +A\rightarrow 2jet +A$. 

Shortest calculation is for the amplitude of high $\kappa_t$ di-jet 
production by pion projectile off the nucleus Coulomb field. This process 
gives negligible contribution at FNAL but not at LHC energy range. The 
conservation of e.m. current is used to express leading term in $s$ in 
terms of the amplitude of transverse photon field -generalization of the 
Weizecker-Williams approximation. Main result is that the dominant 
contribution is given by the amplitude where transverse photon interacts with 
external lines only. This is the  the leading order over $\alpha_s$ 
generalization of the theorems of F.Low and V.Gribov. Thus we proved 
that  high momentum tail of pion wave function of pion is measurable in this 
process. High momentum tail of photon wave function is measurable in the 
similar process: $\gamma+e \rightarrow 2jet +e$ which should be measurable 
at electron colliders(S.Brodsky-comment at this conference).

   We introduce concept of post selection to stress  the fact
that selection of $q\bar q$ component in the final state requires  squizing 
of pion wave function in the initial state because in the average 
configuration around 50\% of pion momentum is carried by gluons. 
Screening of color in the squized meson wf +asymptotic freedom is the only
known effective mechanism of killing of infinite number of degrees of
freedom  characterizing hadron in the average configuration. Sudakov form
factors is well known pQCD analogue of this phenomenon which is also 
important but for the squized initial state pion wave function only .
 
  We calculate imaginary part of amplitude and restore its real part 
by virtue of dispersion relations. In the leading order over $\alpha_s$
and all orders over $\alpha_{s}\ln{k_t^2\over \lambda_{QCD}^2}$ Ward
identities have the same form as in QED. Except impulse
approximation other diagrams correspond to skewed gluon distribution where
the fraction of nucleon momentum carried by gluon in the final state is
negative. To avoid strong suppression this gluon should be wee parton. 
One should also to account for
the cancellation between radiative correction to the wf of final state
and gluon bremstrahlung in the jet direction. All in all lead 
to large virtuality of quarks(similar to that in $e\bar e\rightarrow hadrons$) 
and to the suppression of all diagrams except those 
where color in initial hadron is concentrated in the
transverse region $\propto 1/k_t^2$  . In the evaluation of impulse
approximation infrared divergence are cancelled out after proper account 
of renormalization invariance-renormalization of quark,gluon propagators
and vertexes. In ref.\cite{ref22} it has been suggested to use special l.c.
gauge as the effective method to remove infrared divergencies.
Thus it was shown that high momentum tail of pion wf is measurable in the
considered process. Observed $\kappa_t$,$z$ dependencies of cross section are
in a reasonable agreement with one gluon exchange tail of pion wave
function. Relationship between pion distribution function and pion wave
function made it possible to calculate distribution function at large
virtuality.




\begin{center}
\section*{Light-Cone Wavefunction Representations of
GPD and SSA}
Dae Sung Hwang
\end{center}

\begin{center}
Department of Physics, Sejong University, Seoul 143-747, South Korea
\end{center}


We give a complete representation of virtual Compton scattering
$\gamma^* p \to \gamma p$ at large initial photon virtuality $Q^2$ and
small momentum transfer squared $t$ in terms of the light-cone
wavefunctions of the target proton. We verify the identities between
the skewed parton distributions $H(x,\zeta,t)$ and $E(x,\zeta,t)$
which appear in deeply virtual Compton scattering and the
corresponding integrands of the Dirac and Pauli form factors $F_1(t)$
and $F_2(t)$ and the gravitational form factors $A_{q}(t)$ and
$B_{q}(t)$ for each quark and anti-quark constituent.  We illustrate
the general formalism for the case of deeply virtual Compton
scattering on the quantum fluctuations of a fermion in quantum
electrodynamics at one loop \cite{BHMS,BDH}.

Recent measurements from the HERMES and SMC collaborations show a
remarkably large azimuthal single-spin asymmetries $A_{UL}$ and
$A_{UT}$ of the proton in semi-inclusive pion leptoproduction
$\gamma^*(q) p \to \pi X$.  We show that final-state interactions
from gluon exchange between the outgoing quark and the target
spectator system lead to single-spin asymmetries in deep
inelastic lepton-proton scattering at leading twist in
perturbative QCD; {\em i.e.}, the rescattering corrections are not
power-law suppressed at large photon virtuality $Q^2$ at fixed
$x_{bj}$.  The existence of such single-spin asymmetries requires
a phase difference between two amplitudes coupling the proton
target with $J^z_p = \pm {1\over 2}$ to the same final-state, the
same amplitudes which are necessary to produce a nonzero proton
anomalous magnetic moment.
We show that the complex phase from the final-state interaction
depends on the angular
momentum $L^z$ of the proton's constituents and is thus distinct
for different proton spin amplitudes \cite{BHS1,BHS2}.



\begin{center}
\section*{Quantum power correction to the Newton law}
I.B. Khriplovich, G.G. Kirilin
\end{center}


\begin{center}
Budker Institute of Nuclear Physics, 630090, Novosibirsk, Russia
\end{center}


The one-loop relative quantum correction of first order in $\hbar$
to the Newton law should look by dimensional reasons as
$l_p^2/r^2$, where $l_p= k \hbar/c^3=1.6\cdot 10^{-33}$ cm is the
Planck length. Then, the absolute correction to the potential is:
\begin{equation}\label{vq}
U_{qu}= a_{qu}\,\frac{k^2 \hbar\, m_1 m_2}{c^3
r^3}.
\end{equation}
The problem is to find the numerical constant $a_{qu}$. In spite
of extreme smallness of the quantum correction, its investigation
certainly has a methodological interest: this is a closed
calculation of a higher order effect in the nonrenormalizable
quantum gravity.

The reason why this problem allows for a closed solution is as
follows. The Fourier-transform of $1/r^3$ is
\begin{equation}\label{fur}
\int d{\bf r} \,\frac{\exp (-i {\bf q} {\bf r})}{r^3}\,= -\,2\pi
\ln q^2.
\end{equation}
This singularity in the momentum transfer $q$ means that the
correction discussed can be generated only by diagrams with two
massless particles in the $t$-channel. The number of such diagrams
of second order in $k$ is finite, and their logarithmic part in
$q^2$ can be calculated unambiguously.

We have found the graviton contribution (which is dominant
numerically) to the discussed correction. The result for the thus
modified Newton potential is
\begin{equation}\label{f}
U(r)=-\,\frac{k m_1 m_2}{r}\,\left(1+\,\frac{121}{10\pi}\, \frac{k
\hbar}{c^3 r^2}\right).
\end{equation}

The previous calculations of this contribution to the discussed
effect are demonstrated to be incorrect.



\begin{center}
\section*{Factorization of hard amplitudes in QCD}
L. Mankiewicz
\end{center}


\begin{center}
Center for Theoretical Physics, 
Polish Academy of Sciences \\  
        al. Lotnik\'ow 32/46, Warsaw, Poland, and \\
        Institut f\"ur Theoretische Physik, Universit\"at
Regensburg \\ D-93040 Regensburg, Germany
\end{center}

Were QCD a linear or at least weak coupling theory, our life would
have been much easier. When classical system described by linear
physics is disturbed by an external probe, it responds with excitation
of those degrees of freedom which carry the same frequencies as the
probe itself. Strictly speaking in quantum mechanics it is only true
in the Born approximation as higher orders of perturbation theory
involve communication between degrees of freedom of various
frequencies (summation over the intermediate states) but this does not
lead to any real difficulty as long as weak coupling expansion is
justified. In those two cases one can indeed speak degrees of freedom,
or physics, which plays the leading role a given process.

In QCD the situation is different. The nonlinearity is strong and in a
typical situation, even when the probe carry only high frequencies,
after a few hard interactions the frequency spectrum spreads out into
the low scale, non-perturbative range. The resulting situation is
similar to that of Navier-Stokes equation in the turbulent regime -
the spectrum of relevant degrees of freedom spreads out over the whole
kinematically allowed range, which makes the problem untractable by
any known computational scheme.  It is remarkable one can still define
a class of QCD processes which obey the factorization. Splitting the
amplitude into hard, perturbatively computable, and soft, long-distance
parts allows for set of non-trivial predictions which can be tested
experimentally. More about that can be found in other contributions to
these proceedings. Here we mention only that factorization is born
deeply into the general structure of relativistic field theory - the
requirement of causality in hard kinematics severely restricts the
possibility of interaction between quanta of different frequencies.

As the world around us is built mainly of hadrons one will never
achieve a deeper understanding of the ``standard model'' physics below 1 TeV
scale without understanding the strong interactions. A good example is
the ongoing discussion of the CP violating effects in B-meson decays
where the (mis)under\-stand\-ing of the QCD ``factorization'' makes
testing of the electroweak theory in this system to a precision
required by cosmological implications (baryon asymmetry)
difficult, if not impossible.


\begin{center}
\section*{Chiral Symmetry and Light Front Wave Functions of Nuclei: From $A=\infty$
  to $A=1$.}
Gerald A. Miller
\end{center}
\maketitle


\begin{center}
University of Washington,  Seattle, WA 98195-1560, USA
\end{center}

 The motivation for light front dynamics\cite{one}
and  the salient features of the technique\cite{one} formed the introduction.
The applications concerned
heavy nuclei {\em vs} the EMC effect\cite{two},
         the deuteron\cite{three},
      and     nucleon form factors\cite{four}.
Chiral symmetry does not play a big role
 in 
heavy nuclei, hence the need to include the deuteron. 
Computations of  deuteron form factors  at high
momentum transfer depend on
 nucleon form factors\cite{three}, so these are also discussed\cite{four}.
 
Mean field models of nuclei can not
explain the EMC effect.
Including correlations does
not help. This is because 
the Hugenholtz-van
Hove theorem restricts the total plus momentum carried by nucleons\cite{two}.
Including mesons
can lead to a reproduction of the EMC effect, but then one over-predicts the
content of the nuclear sea, in disagreement with experiment.

Chiral symmetry is
needed to obtain a deuteron bound state\cite{three},
if one uses a relativistic pion-only model.
Rotational invariance
is respected,  even when  not manifest in the
formalism\cite{three}. Using the different
 models of   nucleon form factors
leads to a wide spread in calculated deuteron form factors\cite{three}.

Thus there is a need to use the light front to obtain  nucleon form
factors. Our  model, in which the nucleon is treated
as a relativistic system of three bound constituent
    quarks surrounded by
    a cloud of pions,
    achieves a very good
    description of existing data for the four electromagnetic elastic
    form factors\cite{four}.



\begin{center}
\section*{The soft physics responsible for single spin asymmetries
in hard reactions}
P.J.~Mulders
\end{center}

\mbox{}\\[-1.5cm]
\begin{center}
Department of Theoretical Physics, FEW, Vrije Universiteit\\
De Boelelaan 1081, 1081 HV Amsterdam, the Netherlands
\end{center}


In deep inelastic scattering (DIS) the structure functions can be
expressed in terms of quark distribution functions (DF), which 
have a natural interpretation as quark densities
in the lightcone (plus) momentum fraction $x = p^+/P^+$ (with $p$
being the quark momentum and $P$ the target hadron momentum).
The DF can be written down as lightcone correlation functions of quark
fields with the nonlocality being just in a lightlike (minus)
direction. 
Three DF exist for a spin 1/2 hadron, 
$f_1^q(x) = q(x)$, $g_1^q(x) = \Delta q(x)$ and $f_1^q(x) = 
\delta q(x)$, the latter not accessible in deep-inelastic
scattering. Matrix elements including longitudinal
($A^+$) gluons are absorbed into the definition of the
DF providing the gauge link.

Semi-inclusive DIS involves two hadrons. Including also the hard scale $q$,
one has three momenta, hence besides the lightcone plus and minus
directions, transverse momenta are relevant. One needs to consider
DF and fragmentation functions (FF) depending on transverse
momenta of partons. The functions appear in
azimuthal asymmetries involving the azimuthal angle of the
production plane with respect to the 
lepton scattering plane in leptoproduction.

Single spin asymmetries in hard cross sections can be
traced to T-odd DF and FF~\cite{refmul1}. While T-odd FF, such as the Collins
function describing the fragmentation of transversely
polarized quarks into pions, appear naturally~\cite{refmul2}, 
T-odd DF are less natural. All T-odd effects
at leading order involving hadrons with spin $S < 1$,
require transverse momentum dependent DF and FF.
The operator structure of the
corresponding soft parts requires operators beyond leading
twist, which under specific conditions also allow T-odd effects
for DF~\cite{refmul3}. The latter is connected to the fact that
the gauge link has to bridge transverse separations in the
quark-quark correlation functions, requiring not only longitudinal
but also {\em leading}
transverse gluon contributions.




\begin{center}
\section*{ Instanton vacuum QCD effective action beyond chiral limit}
M.~Musakhanov\end{center}


\begin{center}
Theoretical  Physics Dept, Uzbekistan National University, \\
Tashkent 700174, Uzbekistan, E-mail: musakhanov@nuuz.uzsci.net
\end{center}

Instantons represent a very important  component of the QCD vacuum.
Their properties are described by the average instanton size
$\rho\sim 1/3 \, {\rm fm}$ and inter-instanton distance
$ R\sim 1 \, {\rm fm}$ \cite{shu82}.
The  Lee$\&$Bardeen's fermionic determinant ${\det}_N$ (in the field of
$N_+$ instantons and  $N_-$ antiinstantons) is \cite{lee79}:

\begin{equation}
{\det}_N=\det B, \,\, B_{ij}=
im\delta_{ij} + a_{ji}, \,\,
a_{-+}=-<\Phi_{- , 0} | i\rlap{/}{\partial} |\Phi_{+ , 0} > .
\end{equation}
$\Phi_{\pm , 0} $ are quark zero-modes
generated by instantons(antiinstantons).
${\det}_N$ averaged over
instanton/anti-instanton positions, orientations and sizes
is a partition function of light quarks $Z_N$.
Small packing parameter
$\pi^2 (\frac{\rho}{R})^4 \sim 0.1$  provide 
independent averaging over
instanton collective coordinates.
We calculate  $Z_N$ by the ferminonisation of it \cite{musak99},
which is the natural way to introduce the constituent quarks.
But it is a non-unique procedure beyond chiral limit.
Natural way of resolving this problem is an
application of the axial-anomaly low-energy theorems \cite{musak97}.
Then the partition function satisfying  the theorems is  as follows:
\begin{equation} Z_N = \int D\psi D\psi^\dagger \exp
(\int d^4 x
(\psi^{\dagger}(i\hat\partial \,+\, im)\psi
+\sum_\pm \left(\frac{2V}{N}\right)^{N_f - 1}
(i M)^{N_f}\int \frac {d^4 kd^4 l}{(2\pi )^8}
 \end{equation}
$$\times
 \det_{fg}
 \exp ( -i(k - l)x)
 F(k^2) F(l^2)  \psi^\dagger_f (k) \frac12 (1 \pm \gamma_5 ) \psi_g (l)) .
$$
This effective action , being derived from Lee$\&$Bardeen fermionic
determinant, certainly coincide in the chiral limit with
Diakonov$\&$Petrov effective action \cite{diak86}.
As a first application of this effective action it were calculated
the terms of order$O(m)$ and $O(m^2)$ of Gasser and Leutwyler
phenomenological chiral lagrangian \cite{gass84}
and found that
\begin{equation}
\bar l^{theor}_{3} = 1.39, \,\,
\bar l^{phenom}_{3} = 2.9 \pm 2.4, \,\,
\bar l^{theor}_{4}= 3.57, \,\, \bar l^{phenom}_{4} = 4.3 \pm 0.9.
\end{equation}
We see the good correspondence of the theory with phenomenology
at least for these quantities. The calculations of other couplings
are in the progress.

 I am very grateful to the organizers of the workshop
for the invitation and friendly atmosphere.


\begin{center}
\section*{Hadron Structure and Chiral Physics on the
Lattice}
John W. Negele
\end{center}


\begin{center}
MIT,  77 Massachusetts Ave.,  Cambridge MA
02139, USA
\end{center}


Lattice field theory is the only known way to solve
nonperturbative QCD, and hence plays a crucial role in our
understanding of the structure of hadrons and of the QCD vacuum.
Given the focus of this conference, this talk emphasizes two ways
in which spontaneously broken chiral symmetry is a central issue in
contemporary lattice QCD.
\smallskip

\noindent{\bf Lattice Evidence for the Role of Instantons}

Lattice calculations provide strong evidence for the presence of
instantons in the QCD vacuum, and for their role in producing
spontaneous chiral symmetry breaking and in the physics of light
quarks. Two point correlations of meson currents in the QCD vacuum
show the detailed behavior specified by the instanton-induced 't
Hooft interaction
\cite{Chu:1993cn}.  By minimizing the action locally in a process
known as cooling,  the instanton content of the quenched and full
QCD vacuum was extracted~\cite{Chu:1994vi,in97_1,Negele:1999ev}.
Comparison of hadronic observables calculated with all gluons  and
those obtained using only  the instantons remaining after cooling
yielded qualitative agreement for  hadron masses, quark
distributions, and vacuum correlation functions of hadron
currents~\cite{Chu:1994vi}.  Calculation of the lowest quark
eigenmodes revealed zero modes correlated spatially with the
instantons and truncation of the quark propagators to the zero mode
zone produced the full strength of the $\rho$ and $\pi$
contributions to vacuum correlation functions
\cite {Negele:2000mb}. The topological susceptibility
calculated on the lattice produces the result $(180 MeV)^4$
required by the Veneziano-Witten formula to produce the observed
$\eta^{\prime}$ mass. And finally, using the Banks Casher relation,
the zero modes associated with instantons have been shown
to provide the full value of
the chiral condensate
$\langle \bar\psi \psi \rangle$. Taken together, this whole body of
evidence shows that instantons and their associated quark zero
modes play a central role in light quark physics.
\smallskip

\noindent{\bf Moments of Structure Functions}

Ground state matrix elements specifying
moments of quark density and spin distributions in the nucleon
have been calculated in full QCD\cite{dolgov-thesis,Dolgov:2002zm}.
It has been shown that physical extrapolation of these observables
to the chiral limit  by including  the physics of the pion cloud
resolves previous apparent conflicts with experiment
\cite{Detmold:2001jb} and that computational resources of the order
of 8 Teraflops years are required for a definitive calculation of
the quark structure of the nucleon.




\begin{center}
\section*{Saturation Signals in Hard Diffractive QCD}
N.N.Nikolaev
\end{center}

\begin{center}
IKP FZJ, D-52425 J\"ulich, Germany
\end{center}

The interpretation of nuclear opacity in terms of a fusion
and saturation of nuclear partons has been introduced in
1975 \cite{NZfusion} way before the QCD parton model. The 
pQCD discussion of the fusion has been initiated by by Mueller 
\cite{Mueller1} and 
McLerran and Venugopalan \cite{McLerran}.

 Based on the consistent treatment of intranuclear
distortions, we derive the two-plateau spectrum of 
FS quarks
in the opacity/saturation regime. We
find a substantial nuclear broadening of inclusive FS spectra,
which is especially strong for truly inelastic DIS with color
excitation of the target nucleus, and
demonstrate that despite this broadening the FS sea parton density
exactly equals the IS sea parton density calculated in terms of
the WW glue of the nucleus as defined according to \cite{NSS}. We
pay a special attention to an important point that coherent
diffractive DIS makes precisely 50 per cent of  total DIS 
\cite{NZZdiffr}. We point out that the saturated diffractive
plateau measures precisely the momentum distribution  
in the $q\bar{q}$ Fock state of the photon.
There emerges quite a nontrivial pattern of nuclear effects:
The density of WW glue of an ultrarelativistic Lorentz-contracted 
nucleus per unit area in the impact parameter plane exhibits 
strong nuclear dilution, but because of the 
anti-collinear splitting of WW gluons into sea quarks 
the density of nuclear sea saturates.
The results presented in this talk have partly been published
elsewhere \cite{JETP}.

\begin{center}
\section*{The leading-twist two gluon distribution amplitude 
in exclusive processes involving \\ $\eta$ and $\eta'$ mesons}
Kornelija Passek-Kumeri\v{c}ki
\end{center}
\maketitle


\begin{center}
Fachbereich Physik, Universit\"at Wuppertal, 42097
Wuppertal, Germany
\end{center}


The brief review of work in progress \cite{KrollP02}
has been presented.
The formalism for
treating the leading-twist
two-gluon Fock components appearing in hard processes
that involve $\eta$ and $\eta'$ mesons has been 
reexamined permitting a critical
appraisal of the relevant literature
\cite{Terentev81etc}.
A consistent set of conventions for the definition of
the gluon distribution amplitude, the anomalous dimensions as well as  
the projector of a two-gluon state onto an $\eta$ or $\eta'$ state
has been established.
The $\eta$, $\eta'$--photon transition form factor 
has been  calculated to order $\alpha_s$ 
and as a crucial test of the consistency of this set
of conventions the cancellation of the collinear and UV singularities
has been explicitly shown. 
An estimate of the lowest Gegenbauer coefficients of the
gluon and quark distribution amplitudes has been obtained from a fit to
the $\eta$, $\eta'$--photon transition form factor data \cite{exptff}.
The results were applied to
$\chi \rightarrow \eta \eta (\eta' \eta')$ decays,
deeply virtual electroproduction of
$\eta$ and $\eta'$ mesons (DVEM) and $g^* g^* \eta'$ vertex.
While for DVEM the two-gluon contributions 
are suppressed for small
momentum transfer $t$,
for the $g^* g^* \eta'$ vertex
a significant dependence on $gg$ contributions
has been observed.



\begin{center}
\section*{Review of confinement scenario.}
V.Petrov
\end{center}
\maketitle
The phenomenon of the quark confinement in the Yang-Mills
theories is investigated already for 25 years but up to now no
self-consistent approach which would be able to to explain all known
experimental facts was proposed. It is widely accepted that the
explanation of the quark confinement in the QCD (which implies in turn
the absence of the states with non-integer baryon charge) requires
that the potential between static sources in pure glue theory should
linearly grow with distance. However the way from pure glue theory
to the quark confinement in QCD was never completely traced and there
are some confinement scenario on the market which do not imply linear
potential in the pure glue theory (we mean, first of all, V.N.Gribov's
scenario).

We collect all known theoretical facts about confinement.
These facts lead to the rather contradictory picture of this
phenomenon. From one side confinement is  {\em hard} and based on the
string-like picture (most convincing facts are linear Regge
trajectories and Hagedorne type of all QCD thermodynamics). From the
other side it is rather {\em soft} and almost invisible, as it follows
from the success of perturbation theory (in describing hadron
collisions at high energies), QCD sum rules, different models of the
QCD vacuum (for example, instanton liquid model).

We classify all possible sources of the linear potential in the
Yang-Mills theory. The successful scenario should solve
the puzzle which can be formulated as follows. Linear potential implies
long range correlations in some correlators. From the other side in the
gauge invariant sector all states should be massive and all correlators
should decay exponentially. It is almost imposible to provide both
features simultaneously and, indeed, we demonstrate that all known
scenario of ''elecrical'' confinement lead to massless particles in the
spectrum or even to tachyons.

More promising are scenarios based on the so-called ''magnetic''
confinement. We explain, in details, why (opposite to electrical case)
it is possible to have linear potential and massive spectrum in the
same theory.  We follow all known theoretical examples (dual
superconductor, Polyakov's 2+1 Georgi-Glashow model, SUSY
Witten-Seiberg model) and demonstrate that from the general point of
view they all are based on the same phenomenon in the dual formulation
of the theory.

The main feature of this phenomenon is spontaneous breakdown of the
color group down to $U(1)$ as it was mentioned by t'Hooft many years
ago. In the presented examples this breakdown is introduced in the
theory ''by hands`` but we argue that, in fact, spontaneous brfeakdown
of color is an inherent property of such a confinement mechanism.

We itemize  all possible magnetic objects which can lead to the
confinement in the electrical sector of the theory. First of all one
needs the objects with long ranged interaction: monopoles, vortices and
the object with "$1/r^2$" tail are the only candidates. However these
objects are stable only in the presence of charged condensate
("elementary" or "composite" Higgs effect) and magnetic charge is
quantized also only for this reason. Moreover we demonstrate  that
objects with long-range interaction (e.g. monopoles) cannot survive in
the full  $SU(N)$ environment: if the color in the theory is not broken
these objects induce the dynamical Higgs effect theirselves. The notion
of ''non-Abelian'' monopoles is contradictory --- these objects cannot
exist. All this can be said also about vortices and, in addition,
in the theory with non-broken symmetry they are unstable under decay
into chain of monopole-like objects.
Color can remain unbroken for the dual objects which has $1/r^2$ tail
--- their interaction is weaker but still they are
long-ranged enough in order to provide a linear potential. We discuss
this case in some details and prove that most probably this possibility
leaves  massless particles in the spectrum.

Thus if one would accept  that color group in QCD remains unbroken (we
discuss some possible contradictions with experimental data), one has
not too many possibilities to have linear potential in the pure glue
theory. We do not mention yet another class of scenario which are based
on so-called ''stochastic confinemet`` but, as we show, they are all
relies on the lattice-type mechanism (``compactness'' of the gauge
group) which, surely, cannot be reproduced in the continuum limit.

In other words, at present there is no single satisfactory scenario
of confinement. It is clear only that the explanation of this
phenomenon should come from some unknown dynamics in the
gauge-invariant dual representation of the Yang-Mills theory.




\begin{center}
\section*{Positivity bounds on GPDs}

P.V. Pobylitsa
\end{center}

Generalized parton distributions (GPDs) appear in the QCD description of
various hard processes e.g. deeply virtual Compton scattering and hard
exclusive meson production. GPDs are defined in terms of matrix elements of
parton fields $\phi _{\alpha }\left( x\right) $ over hadron states $%
|P,i\rangle $ (with $x$ being parton momentum fraction)%
\begin{equation}
G_{P_{1}i_{1},P_{2}i_{2}}^{\alpha _{1}\alpha _{2}}(x_{1},x_{2})=\langle
P_{2},i_{2}|\phi _{\alpha _{2}}^{\dagger }\left( x_{2}\right) \phi _{\alpha
_{1}}\left( x_{1}\right) |P_{1},i_{1}\rangle   \label{G-def}
\end{equation}%
The positivity of the norm in the Hilbert space of states 
\begin{equation}
\left\| \sum\limits_{i\alpha }\int \frac{dP^{+}d^{2}P^{\perp }dx}{%
2P^{+}(2\pi )^{3}}\theta (x)f_{i\alpha }(x,P)\phi _{\alpha }\left( x\right)
|P,i\rangle \right\| ^{2}\geq 0  \label{starting-inequality}
\end{equation}%
allows to derive rather nontrivial inequalities for GPDs. Various positivity
bounds for GPDs corresponding to specific choices of the function $%
f_{i\alpha }(x,P)$ have been already discussed in literature \cite{prev}. In
ref. \cite{Pobylitsa-02b} inequality (\ref{starting-inequality}) is analyzed
for arbitrary functions $f_{i\alpha }(x,P)$. As a result new positivity
bounds are derived for generalized parton distributions using
the impact parameter representation. These inequalities are stable under the
evolution to higher normalization points. The full set of inequalities is
infinite.


\begin{center}
{\large\bf Soft pion theorems for partons}\\[0.5cm]
P.V. Pobylitsa$^{1,2}$, {\bf M. V. Polyakov}$^{1,2}$, M.
Strikman$^{1,3}$\\[0.3cm]
$^1$Petersburg Nuclear Physics Institute, Gatchina, 188350 Russia,\\[0.3cm]
$^2$Institute for Theoretical Physics II, Ruhr University,
44780 Bochum,Germany\\[0.3cm]
$^3$ Penn State University, USA\\[0.3cm]
\end{center}

We show that for the
processes $\gamma^*(q) N(p)\to \pi(k) N(p')$ near the
threshold\footnote{
$M_{\pi N}-(M_N+m_\pi) \sim m_\pi$ where $M_{\pi M}$ is the invariant mass of the
produced $\pi$ and nucleon} the two
limits $Q^2\to \infty$ ($Q^2=-q^2$) and $m_\pi\to 0$ do not commute.
For $Q^2\ll \Lambda^3/m_\pi$ ($\Lambda$ is a typical hadronic scale)
one can apply classical soft pion theorems by Nambu~{\em et al.}
\cite{Nambu} which express the threshold amplitudes of the $\gamma^*N\to \pi N'$
reaction in terms of e.m. and axial form factors of the nucleon.
In the kinematics where $Q^2\gg \Lambda^3/m_\pi$ one can derive
new soft pion theorems \cite{Pobylitsa:2001cz} which also express
the threshold amplitudes in terms of the nucleon form factors.
The particular expressions depend on the form of the light-cone wave
functions (LCWF) of the nucleon, $i.e.$ sensitive to the partonic content
of the nucleon wave function.
If we assume that the dominant
component of the nucleon LCWF is symmetric we obtain, for
instance, at $ Q^2\gg \Lambda^3/m_\pi$ we have \cite{Pobylitsa:2001cz}
\begin{eqnarray}
\nonumber
A(\gamma^*p\to \pi^0 p)|_{{\rm th}}&=&
\frac{1}{3 f_\pi}
\left(\frac 5 2\ G_{Mp}(q^2)-4 \ G_{Mn}(q^2)
\right)+O\left(\frac{m_\pi}{\Lambda} \right)
\, ,
\end{eqnarray}
which should be contrasted with \cite{Nambu}
\begin{eqnarray}
\nonumber
A(\gamma^*p\to \pi^0 p)|_{{\rm th}}&=&
\frac{g_A}{ 2 f_\pi}\
G_{Mp}(q^2) +O\left(\frac{m_\pi}{\Lambda} \right)
\, ,
\end{eqnarray}
for $\Lambda^2 \ll Q^2\ll \Lambda^3/m_\pi$.
The new soft pion theorems for hard processes are in agreement
with measurements of $\gamma^*N\to \pi N'$ reaction near threshold
by E136 collaboration \cite{Bosted:1993cc}.
We also discussed soft pion theorems for other hard exclusive
processes, see details in \cite{Polyakov:1998ze}.


\begin{center}
\section*{Pion properties \\ in the non-local chiral quark model}
M. Prasza{\l}owicz
\end{center}


\begin{center}
Particle Theory Dept., M. Smoluchowski Institute of Physics, \\
Jagellonian University, PL-30-059 Krak{\'o}w, Poland
\end{center}


In this talk \cite{mp:bh2002}, following our previous work 
\cite{Praszalowicz:2001wy,mp:talks}
and Refs.\cite{PetPob,Bochum}, we present a simple, nonperturbative model
in which quarks acquire a momentum dependent dynamical mass. The model is based
on the instanton model of the QCD vacuum, however, the form of the nonlocality
is simplified in order to perform all calculations directly in the Minkowski
space-time.

Using this model we calculate pion light cone wave functions and
distribution amplitudes, two pion distribution amplitudes and
skewed parton distributions which allow us to calculate pion
structure function and the electromagnetic form factor. The aim of
this study is twofold. Firstly, we calculate in the unique
framework different characteristics of the pion and secondly, we
perform various tests in order to gain confidence in the model as
well as to find its limitations. We argue that the model,
given its simplicity both conceptual and technical, 
works much better than one might have initially expected.

This work was partially supported by the Polish KBN Grant
PB~2~P03B~{\-} 019~17.






\begin{center}
\section*{Power-Law  Wave Functions and 
Generalized Parton Distributions  for Pion}
A.V.~Radyushkin
\end{center}

\begin{center}
Physics Department, Old Dominion University, Norfolk, VA 23529,
USA \\
Theory Group, Jefferson Lab, Newport News, VA 23606, USA
\end{center}


In Ref. \cite{Mukherjee:2002gb}, we study a model 
based on  a  power-law ansatz 
$\psi(x,k_\perp)\sim 1/{\sqrt{x(1-x)}[a+bk_\perp^2]^2}$ for 
the pion light cone  wave function,  
where  $a=1+s^2 (x-1/ 2)^2/{x(1-x)}$, $ b={1/{4 \lambda^2
x(1-x)}}$; with  $s={m / \lambda}$ being the ratio 
of the effective quark mass $m$  and the 
pion size parameter $\lambda$.
We  derive a parametric expression 
for the pion form factor and  
show that  $m$ and $\lambda$
can be easily adjusted to provide a curve 
close to existing experimental 
data. 

Our  parametric representation for the form factor 
has the form of a reduction relation 
connecting the pion form factor and the double distribution  (DD)
$F(x,y;t)$.  
 This model DD   has the correct spectral and symmetry
properties and it also has the factorized structure
used in phenomenological models:  it looks like  
a distribution amplitude with respect to 
the $y$ variable and like a parton density
with respect to the $x$ variable. 
Furthermore, it provides a nontrivial example of the interplay between
$x$, $y$ and $t$ dependence
 of DDs. With an explicit model for DDs at hand, one can calculate 
 the relevant skewed distributions.

The simple toy model  is not very realistic: the valence parton density
  obtained in this model differs rather strongly
  from the phenomenologically established form.
  We fix this deficiency 
  by adopting a model with a more realistic 
  $x$-profile at $t=0$, but preserving the 
  analytic structure of the interplay 
  between $x,y$ and $t$ dependence generated 
  by the power-law ansatz.
  We show that by slightly changing 
 $m$ and $\lambda$
  it is still possible to get a good description of 
  the pion form factor data. 
  We present skewed parton distributions  (SPDs)  obtained from the 
  ``realistic'' DD. These SPDs satisfy such  important constraints
  as reduction relations to usual parton densities and form factors,
  they also have correct spectral and 
  polynomiality properties, thus providing
  a  model for the pion GPDs
   that can be used in phenomenological
  applications.



\begin{center}
\section*{How quantum mechanics of the glue \\ can help us understand RHIC
puzzles}
E.V.Shuryak
\end{center}
\maketitle


\begin{center}
Department of Physics and Astronomy,
University at Stony Brook, NY 11794 USA
\end{center}


 I was given two talks, on a request of the organizers.
The first summarized
recent data from the Relativistic Heavy Ion Collider at Brookhaven
have displayed a number of striking features, some predicted some not.
The existence of strong collective flow suggests Bang-like
(hydrodynamical) behavior, with very rapid equilibration and
short mean free paths.  The equation of state is close to that predicted
by lattice QCD, suggesting that the experiments really are producing
Quark-Gluon Plasma for the first time since the Big Bang.
Strong jet quenching, the large azimuthal
asymmetry for hadrons with large transverse momentum, and the
apparent absence of backward jets, all point directly to
very strong absorption in the system. Dynamical models
based on perturbative QCD extrapolated down to 1-2 GeV momenta fail
miserably to explain these data quantitatively. At the moment no
non-perturbative explanations of the dynamics are yet available.
The week after Bad Honnef workshop, at Quark Matte 2002 in Nantes,
many of these statements got much stronger support from second RHIC run.

The second talk was about new development in semi-classical theory
of high energy collisions, hadronic or heavy ions
\cite{Shuryak:2002an}  .
Sudden deposition of energy
 makes
virtual  gluon fields real, without significant change
in any of the  coordinates. The same is true for the
topological coordinate, the Chern-Simons number $N_{CS}$.
It implies that virtual  fields under the barrier in QCD vacuum
(instantons) can be excited to states $at$ the barrier. Those are,
  gluomagnetic clusters
of particular structure,  the
{\em ``turning states''} related to electroweak sphalerons and have
 the mass peaked in the range of 2.5-3 GeV.
These clusters
immediately explode into
a {\em thin shell of  coherent transverse field} which then
becomes several outgoing gluons and quarks.
New classical solutions for YM and Dirac eqns are
found describing both processes.
We argue that such clusters
should be multiply produced in heavy ion collisions, and give some
estimates of their effect  at the RHIC energies.






\begin{center}
\section*{Baryon Form Factors and GPD's}
\end{center}

\begin{center}
Paul Stoler
\end{center}

\begin{center}
{\em Physics Department, Rensselaer Polytechnic Institute, Troy, NY 12180}
\end{center}

An important part of the Jefferson Lab (JLab) current and long range programme
is to measure form factors and exclusive reactions out to as high momentum transfer
as possible. In parallel, during the past several years there has been important progress
in the characterization of these reactions at experimentally accessible momentum transfers,
in terms of generalized parton distributions (GPDs). A variety of exclusive
{\it form factor like} reactions provide different moments $< x^m >$ of these GPD's.
For example the elastic form factors give us the $m = 0$ moments, while Compton scattering
and high-$t$ meson production provide the $m = -1$ moments of the $H$ and $E$ GPD's.
The $N \to \Delta$ resonance transition magnetic form factor $G^*_M$ selects
the {\it isovector} components. As a function of $t$ these reactions provide
important constraints on models of the transverse and longitudinal structure of the nucleon
and it's excitations, which were accessible to inclusive deep inelastic scattering (DIS).

As an example, following ref. \cite{Rad} I have obtained GPD's from two-body model wave
functions \cite{Sto} with a wave function having a Gaussian plus power law dependence
on parton $k_\perp$, which is constrained to agree simultaneously with experimental data
on $G_{Mp}$ and $G_{Ep}/G_M$ over the full range of measured $Q^2$. Reasonable results
can then also be obtained for other form factors, such as the $G^*_M$, for the $N \to \Delta$
transition, for which the relevant GPD is approximated by the isovector part of elastic GPD.
This is shown in Figure 1 below. The GPD's are currently being applied
to the neutron form factors using isospin invariance.

\begin{figure}[h!]
  \hspace{-0.8cm}
  \epsfig{figure
  =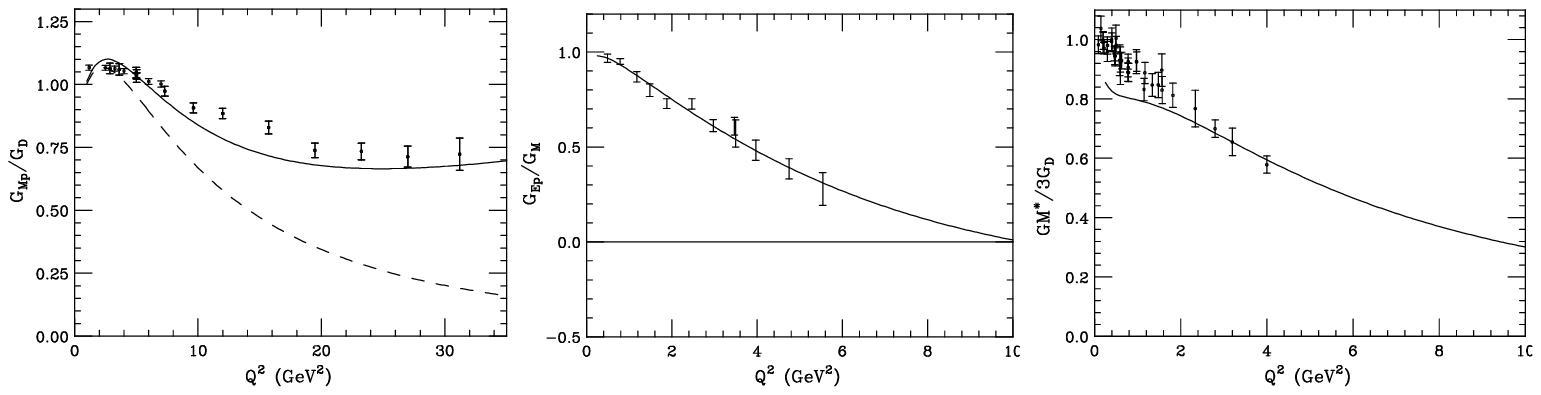,width=13.5cm,height=4cm}
\end{figure}

Figure 1:
Left:
Proton magnetic form factor $G_{Mp}/G_D$, where $G_D = 1/(1+Q^2/0.71)^2$.
The data are from SLAC \cite{Arn}. The curves are GPD fits as in ref. \cite{Sto}.
The dashed curve uses a Gaussian parton $k_\perp$ wave function,
while the solid curve uses a Gaussian plus power $k_\perp$ wave function \cite{Sto}.
Centre:
Proton electric form factor $G_{Ep}/G_M$. The data are from JLab \cite{Gay},
and the curve is the GPD fit.
Right:
The $P \to \Delta$ form factor $G^*_M/3G_D$. The data is a world compilation.
The curve utilizes the {\it isovector} part of the GPD's obtained in the right
and centre elastic form factors, renormalized to account for the measured value
at $Q^2=0$.

In summary, it is expected that the theoretical community will continue
to more rigorously progress with the application of GPDs to high $t$ exclusive reactions
in parallel with the anticipated progress in experimental measurements.


\date{}
\begin{center}
\section*{Small x phenomena and  gluon tomography of nucleon}
M.Strikman
\end{center}


\begin{center}
Penn State University, PA 16802, USA
\end{center}

It was demonstrated in \cite{BFGMS} for the case of small x and in \cite{CFS}
 in a general case 
 that in the 
limit of large $Q^2$ the amplitude of the 
process $\gamma^*_L + N \to V + N$ at fixed $x$ 
is factorized into the convolution of a hard interaction block
calculable in perturbative QCD,  the  short-distance 
$q\bar q $ wave function of the meson, and the generalized/skewed
parton distribution (GPD) in the nucleon.
 The t-dependence of GPD's provides unique information about the
impact parameter distribution of the partons in nucleons.
The data have confirmed a number of the predictions of \cite{BFGMS}
including approximate  restoration of SU(3) symmetry, convergence of the 
t-slope of the $\rho-$meson production at large $Q^2$ to the slope
of the $J/\psi $ production. 
The extension of 
the analysis of  \cite{BFGMS} in \cite{FKS96} to account for the 
finite transverse size effects for 
$J/\psi$ production indicates the squeezing  starts  already 
from $Q^2\sim 0$. The difference of the 
t-dependences of  $\rho$ and $J/\psi$ 
 predicted  in \cite{FKS96}  agrees well 
with the recent HERA data.
This allowed us \cite{FS2002} to use the data on $J/\psi$ photoproduction
to extract the two-gluon form factor  of nucleon
$\Gamma(t)$ from the data. We find that 
$\Gamma(t)=(1-t/m_{2g}^2)^{-2}$ with 
$m_{2g}^2\approx 1 GeV^2$
provides a good description of the $t$-dependence of
the cross section of the elastic                
 photoproduction of $J/\psi$-mesons 
between the threshold region of $E_{\gamma}=11~ GeV$ (Cornell), 
$E_{\gamma}=19~ GeV$ (SLAC)
  and  $E_{\gamma}=100~ GeV$ (FNAL) including the
 strong energy dependence of the t-slope.
It is also well matched with the recent HERA data.
We argue that a larger mass scale entering in $\Gamma(x_i\ge 0.05,t)$ 
than in the e.m. form factor 
is due to suppression of scattering off the gluons belonging
the soft pion fields for $x\ge 0.05$.





\begin{center}
{\Large \bf Deeply virtual Compton scattering with soft pion production}
\end{center}

\vspace*{6mm}


\large
\begin{center}
{Marc Vanderhaeghen }\\
\end{center}

\begin{center}
\small{\it Institut f\"ur Kernphysik, University Mainz, D-55099 Mainz, Germany}
\end{center}

\vspace*{10mm}

\normalsize
\noindent 
We discuss the extension of the Generalized Parton Distribution (GPD) 
formalism to hard exclusive processes 
such as deeply virtual Compton scattering (DVCS)
involving the production of a soft pion. The evaluation of this
process is relevant for DVCS experiments which do not have the resolution
to distinguish the nucleon final state from a nucleon in addition with
a soft pion. 
\newline
\indent
In addition, the DVCS processes with an additional soft pion is also
of interest by itself. 
Indeed, by using soft-pion theorems, it is shown to be possible to link 
the GPDs entering in the ordinary DVCS process and the DVCS process with an
associated soft pion produced \cite{Ref-1}. 
\newline
\indent
We also give first numerical predictions for the DVCS processes on the
nucleon with an additional soft $\pi^0$ or $\pi^+$ and compare it with the
original DVCS process.


\begin{center}
\section*{Experimental Tests of Chiral Dynamics}
Thomas Walcher
\end{center}


\begin{center}
Institut f\"ur Kernphysik\\
Johannes-Gutenberg Universit\"at Mainz\\
Mainz, Germany
\end{center}


The contribution discussed three different aspects of chiral dynamics.  
In the first part some recent experiments close to the $\pi N$ 
threshold were presented showing good agreement for experiments 
with photons $(Q^2 = 0)$ \cite{asch01} but rather severe deviations for 
experiments with electrons $(Q^2 \leq 0.1 (GeV/c)^2)$ \cite{hme02}. 
On the other hand, the generalized polarizabilities determined in virtual
Compton scattering $p(e,e´p)\gamma$ at $0.3~(GeV/c)^2$ agree well with
the calculations of chiral perturbation theory \cite{ndh01}.
This means that one may have to consider not so much a ``convergence
radius'' of chiral perturbation theory, but some additional dynamics 
of the $\pi$ cloud. 

The second part took up this idea and 
showed how the observed narrow excited states of the nucleon below the 
$\pi N$ threshold could be explained in the framework of chiral
dynamics \cite{thw01}. 

Thirdly, an attempt was made to put the ideas of 
effective field theories in the context of general many body physics. 


%
%
%
%
%
%
\begin{center}  
\begin{large}  
{\bf QCD operators and the chiral Lagrangian}
\end{large}  
\\[.2cm]
{\bf C. Weiss}
\\[.2cm]
Institut f\"ur Theoretische Physik, Universit\"at Regensburg,
D--93053 Regensburg, Germany
\end{center}
The low--energy behavior of strong interactions is described by the chiral 
Lagrangian, formulated as an expansion in derivatives of the pion field.
It allows one to compute not only scattering amplitudes 
of soft pions, but also matrix elements of the vector and axial vector 
current operators between multipion states, describing electroweak 
interactions of pions. However, these are by no means the only
``interesting'' operators one can evaluate between
soft--pion states. Certain classes of composite QCD operators
arise in the factorized description of hard scattering processes
in QCD, such as deeply virtual Compton scattering or 
hard electroproduction of multipion states, whose matrix elements define 
generalized parton distributions (GPD's) or distribution amplitudes 
of the pions. These operators can be normalized at a low scale 
$\mu^2 < 1 \, {\rm GeV}^2$ (independent of the hard scale $Q^2$ of the 
scattering process), and the study of their matrix elements between 
soft-pion states is a legitimate problem of low-energy physics. 
These matrix elements are intesting in themselves, 
as well as as input for the calculation of chiral
loop corrections to both pion and nucleon matrix elements.

In this talk I review some aspects of the problem of ``bosonization''
({\it i.e.}, translation into pion operators) of QCD operators 
appearing in the factorized description of hard
scattering processes in QCD. While the vector and axial vector current 
operators of the effective chiral theory could be identified on grounds of 
their symmetry properties alone, this is generally not possible for the 
QCD operators appearing in the factorization of hard 
processes. A non-trivial exception is the energy--momentum tensor (EMT) of 
QCD, which is the Noether current related to translational invariance.
Identifying the traceless part of the QCD EMT with the one derived from the 
chiral Lagrangian one obtains a sum rule for the second moment of the 
pion GPD's [1]. A similar reasoning applied to the trace anomaly allows
to determine the long--range interaction of heavy quarkonia from first 
principles [2].

In general, the translation of QCD composite operators into pion 
operators requires dynamical information. A very useful model in this 
respect is the picture of the QCD vacuum as ``medium'' of instantons [3]. 
It allows not only to derive the chiral Lagrangian ({\it i.e.}, the constants 
$F_\pi^2$, {\it etc.}) but also provides a script for the bosonization of QCD 
operators, including such containing gluon fields [4]. This approach 
preserves subtle structures such as relations following from the QCD 
equations of motion or the $U(1)$ and trace anomalies. In particular, we \
review the predictions for higher--twist matrix
elements (quark--gluon correlations) determining power corrections 
to deep--inelastic scattering.

\noindent
[1] M.~V.~Polyakov and C.~Weiss, Phys.\ Rev.\  {\bf D60} (1999) 114017.
\\[0cm]
[2] H.~Fujii and D.~Kharzeev,
Phys.\ Rev.\ D {\bf 60}, (1999) 114039.
\\[0cm]
[3] D.~Diakonov and V.~Petrov, Nucl.\ Phys.\  {\bf B245} (1984) 259;
{\bf B272} (1986) 457. 
\\[0cm]
[4] D.~I.~Diakonov, M.~V.~Polyakov and C.~Weiss,
Nucl.\ Phys.\  {\bf B461} (1996) 539.


%
%
%
%

\end{document}